\documentclass{ws-ijmpd}

\begin{document}

\markboth{A.V. Khugaev and B.J. Ahmedov}
{ Remarks on Papapetrou
Solutions of Einstein Equations}


%
\catchline{}{}{}{}{}
%

\title{Remarks on Papapetrou Class of \\Vacuum
Solutions of Einstein Equations }

\author{\footnotesize A.V. KHUGAEV}

\address{Institute of Nuclear Physics, Ulughbek, Tashkent 702132, Uzbekistan\\
Inter University Centre of Astronomy and Astrophysics\\ Post Bag
4, Ganeshkhind 411007, Pune, India\\ avaskhugaev@yahoo.com }

\author{B.J. AHMEDOV}

\address{Institute of Nuclear Physics,
Ulughbek,
         Tashkent 702132, Uzbekistan         \\
Inter University Centre of Astronomy and Astrophysics\\ Post Bag
4, Ganeshkhind 411007, Pune, India\\
    Ulugh Beg Astronomical Institute, Astronomicheskaya 33,
    Tashkent 700052, Uzbekistan\\
    ahmedov@astrin.uzsci.net
}

\maketitle

\begin{history}
\received{15 January 2004}
\end{history}


\begin{abstract}
Class of axially symmetric solutions of vacuum Einstein field
equations including the Papapetrou solution as particular case has
been found. It has been shown that the derived solution describes
the external axial symmetric  gravitational field of the source
with nonvanishing mass. The general solution is obtained for this
class of functions. As an example of physical application, the
spacetime metric outside a line gravitomagnetic monopole has been
obtained from Papapetrou solution of vacuum equations of
gravitational field.

\end{abstract}



\section{Introduction}  
\label{intr}

A wide range of massive objects from the compact relativistic
stars, as black holes and neutron stars,  up to modern
cosmological models, including Early Universe and its
evolution~\cite{HawPen} where the solutions of Einstein equations
are applicable and important is currently known. But the logic of
scientific development and practical experience requires an
introduction into the consideration of principally new objects,
like cosmic strings, gravitomagnetic monopoles, quasiclosed worlds
etc. The extension of objects under research leads to the
additional analysis and extension of well known existing solutions
of fundamental field equations. Therefore reconsideration of old
and finding new exact exterior solutions of Einstein field
equations describing space-time around these objects is one of the
interesting and important branches of general relativity and
theoretical physics. There exist a lot of interesting problems,
especially in quantum gravity, because it is still not clear how
to build quantum gravity (see, for example,~\cite{Dadhich}) and in
this level of our understanding it is very useful to investigate
objects which are on the boundary between classical and quantum
gravity. Of special importance also are external gravitational
fields of isolated stars, as today is recognized they are at the
core of most intriguing astrophysical phenomena. In spite of the
fact that many exact solutions of Einstein equations can be found
in classical monographs as~\cite{Kramer} it is still interesting
to study new possible solutions of this fundamental equation.

In our present work we concentrate our efforts in investigation
and analysis of some exact solutions of Einstein
equations~\cite{Kramer}, namely, on the exterior axisymmetric
fields in the vacuum, arising around rotating gravitational
objects. In our consideration we look for some mathematical
features and structure of Einstein equations, which could
determine their physically acceptable solutions.

 In fact, our motivation is, that the Papapetrou  class of
axially-symmetric solutions~\cite{Papa}$^{,}$\cite{Djamal} of
vacuum Einstein equations does not contain physically
 interesting one, because in the classical
Newtonian limit there
 is no term being proportional to $r^{-1}$, what means
that it describes
 rotation field for the bounded systems with zero mass~\cite{Djamal}.
However it should be noted, that one of the solutions of Einstein
equations with the similar asymptotic behavior has been considered
and interpreted as gravitational field of massless
quadrupole~\cite{Fomin1}. From geometrical point of view, the
gravitational source is characterized  by a topological closure of
internal three dimensional space with zero integral contribution
according to its zero mass.

 In the Papapetrou case the solutions are given
in the class of pure
 harmonic
functions $\zeta$ and our assumption is to
investigate this
 solution in the
more extended functional space of functions $\zeta$.

 Here we follow to the logic of investigation, brilliantly
reflected
in~\cite{Djamal}. As we will show in the Section~\ref{ext}, small
 difference, introduced by us into the approach described in~\cite{Djamal},
 gives us a possibility to obtain solution of Einstein equations
 for more general class, which includes Papapetrou one
 as particular case.

 Throughout, we use a space-like signature $(+,-,-,-)$ and a system
 of units in which $G = 1 = c$.
Greek indices are taken to run
from 0 to 3 and
Latin indices from 1 to 3; partial derivatives are
denoted with a
comma.

\section{Extension of Papapetrou class of solutions}
\label{ext}

We consider a stationary axially symmetric object in cylindrical
coordinates $x^\alpha =(t,\rho,\phi, z)$ and assume that there is
no any gravitating matter outside of it.  For the
rotating fields
in GR, the most general form of the metric of
space-time of axially
symmetric objects can be written as Weyl -
Lewis - Papapetrou one,
presented in the form~\cite{Djamal}
\begin{equation}
ds^2=f(dt-\omega d\phi)^2-\rho^2\cdot f^{-1}d\phi^2
-e^{\mu}(d\rho^2+dz^2)\ ,
\end{equation}
where $fl+k^2=\rho^2$, $f, \omega, \mu, k, l$ are unknown metric
functions may depend on the cylindrical coordinates $\rho$
and $z$.

The external gravitational field is described by the symmetric
Ricci tensor $R_{\alpha\beta}$ which obeys to  the exterior
Einstein vacuum equations $R_{\alpha\beta}=0$.  The nonvanishing
$R_{00}$ and $R_{03}$ components of Einstein equations can be written
as~\cite{Djamal}
\begin{eqnarray}
\label{2}
&&f(f_{,\rho\rho}+f_{,zz}+\rho^{-1}f_{,\rho})-f^2_{,\rho}-
f^2_{,z}+
\rho^{-2}f^{4}(\omega^2_{,\rho}+\omega^2_{,z})=0\ ,\\
\nonumber\\
\label{3}
&&f(\omega_{,\rho\rho}+\omega_{,zz}-\rho^{-1}\omega_{,\rho})
+2f_{,\rho}\omega_{,\rho}+2f_{,z}\omega_{,z}=0\ .
\end{eqnarray}
 Following to the approach presented in~\cite{Djamal} one can select the
 function $\omega$ satisfying to equation
\begin{equation}
\label{4}
\omega_{,\rho\rho}+\omega_{,zz}-\rho^{-1}\omega_{,\rho}=0\ ,
\end{equation}
as $\omega=C\rho\zeta_{,\rho}$, where $C$ is a constant.
Mathematical properties of function $\zeta$ will be defined below.
It is easy to see, that after substituting $\omega$ in (4) and
using simple relations
\begin{eqnarray}
\label{5} &&\omega_{,\rho}=C(\zeta_{,\rho}+\rho\zeta_{,\rho\rho}),
\quad
\omega_{,\rho\rho}=C(2\zeta_{,\rho\rho}+\rho\zeta_{,\rho\rho\rho})\
,
\nonumber \\ \nonumber\\
&&\omega_{,z}=C\rho\zeta_{,\rho z}, \quad
\omega_{,zz}=C\rho\zeta_{,\rho zz}\ ,
\end{eqnarray}
one can write
\begin{equation}
\label{zeta} \zeta_{,\rho\rho\rho}+ \zeta_{,\rho zz}
+(\rho^{-1}\zeta_{,\rho})_{,\rho}=0\ .
\end{equation}
Integration of the equation~(\ref{zeta}) gives an expression for
$\zeta$ :
\begin{equation}
\label{pois}
\nabla^2\zeta=\zeta_{,\rho\rho}+
\zeta_{,zz}+\rho^{-1}\zeta_{,\rho}=\eta (z)\ .
\end{equation}

Therefore we introduced an arbitrary function $\eta=\eta (z)$.
Introduction of this function is very important, because it could
generate different kind of solutions for set of differential
 equations (\ref{2}) and (\ref{3}).
Hereafter we have essential difference with compare
to~\cite{Djamal} and our aim is to solve Einstein equations
(\ref{2}) and (\ref{3}) in terms of more wide class of functions
$\zeta$, which include harmonic functions as a particular case.

In the case $\eta (z)=0$ equation (\ref{pois}) becomes Laplace one, which is
similar to the expression given in~\cite{Djamal}.
However in the general case when $\eta (z)\not=0$,
we have Poisson like equation (\ref{pois})
for function $\zeta$ instead of Laplace one. Using the relations (\ref{5})
for the function
$\omega$ and condition (\ref{4}) one can rewrite field equation
(\ref{3}) as
\begin{equation}
\label{3_mod}
-f_{,\rho}(\zeta_{,zz}-\eta)+f_{,z}\zeta_{,\rho z}=0 \ .
\end{equation}

It follows from the equations (\ref{pois}) and (\ref{3_mod})
the general solution for the metric function $f$ takes
the following form
\begin{equation}
f=f\left(\zeta_{,z}-\int\eta (z)dz\right)\ .
\end{equation}

After introducing new variable $\tilde\zeta_{,z}$ and derivative
$f'$
\begin{equation}
\label{tilzetd}
\tilde\zeta_{,z}=\zeta_{,z}-\int\eta (z)dz, \quad
f'=\frac{\partial f}{\partial\tilde\zeta_{,z} }
\end{equation}
one can rewrite the derivatives of the metric function $f$
in the following way
\begin{eqnarray}
\label{deriv}
&& f_{,\rho}=f'\zeta_{,z\rho},\quad f_{,\rho\rho} =
f''\zeta^2_{,z\rho}+f'\zeta_{,z\rho\rho}\ ,\nonumber
\\ \nonumber\\
&& f_{,z}=f'\left[\zeta_{,zz}-\eta(z)\right],\quad
f_{,zz}=f''(\zeta_{,zz}-\eta )^2+ f'(\zeta_{,zzz}-\eta_{,z})\ .
\end{eqnarray}

Using the equations (\ref{pois}) and (\ref{deriv}) one can get
\begin{eqnarray}
\label{12}
&&f_{,\rho\rho}+f_{,zz}+\rho^{-1}f_{,\rho}=
f''\left[\zeta^2_{,z\rho}+(\zeta_{,zz}-\eta)^2\right]\ ,\\
\nonumber\\
\label{13}
&& f^2_{,\rho}+f^2_{,z}=
f'\left[\zeta^2_{,z\rho}+(\zeta_{,zz}-\eta)^2\right]\ ,\\
\nonumber\\
\label{14} && \omega^2_{,\rho}+\omega^2_{,z}=
C^2\rho^2\left[\zeta^2_{,z\rho}+(\zeta_{,zz}-\eta)^2\right]\ .
\end{eqnarray}

Finally, after inserting the derived expressions into the equation
(\ref{2}) we obtain equation
\begin{equation}
f f''-f'^2+C^2f^4=0
\end{equation}
which looks like an extension of Papapetrou class of solutions of
gravitational field equations:
\begin{equation}
\label{16} f(\rho , z)=\left\{\alpha\cdot
\cosh\left[\zeta_{,z}-\int\eta (z)dz\right] -\beta
\sinh\left[\zeta_{,z}-\int\eta (z)dz\right]\right\}^{-1}\ ,
\end{equation}
where the functions $\eta(z)$ and $\zeta (\rho , z)$  satisfy to the equation
(\ref{pois}) and $C^2=\alpha^2-\beta^2$. Here parameters
$\alpha$ and $\beta$ have the same meaning as in the Papapetrou
solutions. As a particular case, in the limit $\eta (z)=0$, the
solution (16) reduces to the Papapetrou one~\cite{Papa}$^{,}$\cite{Djamal}.

\section{Asymptotic properties}

As a rule, one of the main difficulties, after finding an exact
solution of Einstein equations
 is a physical interpretation of it.  In the case of
 the Papapetrou solutions  in the Newtonian limit that is at large
arguments $r=\sqrt{\rho^2+z^2}\to\infty$,
the asymptotics looks like
\begin{equation}
\label{17} f(\rho , z)=\alpha^{-1}\left[1+\frac{\beta z}{\alpha
r^3}+O(r^{-2})\right]
\end{equation}
and does not contain a term being proportional to $r^{-1}$. From
comparison of expression (\ref{17}) with the metric given in~\cite{MTW}
\begin{eqnarray}
ds^2=\left[1-\frac{2M}{r}+O(r^{-2})\right]dt^2-
\left[4\varepsilon_{ijk}\frac{S^jx^k}{r^3}+
O(r^{-3})\right]dt\cdot dx^i- \nonumber \\
\nonumber\\
  -
\left[(1+\frac{2M}{r})\delta_{jk}+O(r^{-1})\right]dx^jdx^k\ ,
\end{eqnarray}
 one can
conclude that for any given harmonic function $\zeta$, the metric
(\ref{16}) with $\eta (z)=0$ does not contain solution with
nonzero mass of the rotating axially symmetric
source~\cite{Djamal}. Here $M$ is the mass of the source and
the three dimensional
spatial  vector $S^{j}$ is the total angular momentum.

We expect that in general case when $\eta\not= 0$  it would be possible to
construct a solution, which gives flat asymptotics at infinity for
massive sources. As an example, one of the possible solutions
could be obtained in the following way. Assume that the function
$\zeta (\rho ,
z) = f_1(\rho )f_2(z)$ could be taken as a product of two functions
$f_{1}$ and $f_{2}$. Then from equation~(\ref{pois}) one can get
\begin{equation}
f_{1,\rho\rho}f_2+f_1 f_{2,zz}+\rho^{-1}f_{1,\rho}f_2=\eta (z)\ .
\end{equation}

Taking solution for the function $f_2$ in the form
$f_2=f_{2,zz}=\eta (z)$, with
 $\eta(z)=C_0 (e^z-e^{-z})$, we have for the function
 $f_1$ the following equation
 \begin{equation}
 g_{1,\rho\rho}+\rho^{-1}\cdot g_{1,\rho}+g_1=0\ .
 \end{equation}
 Here function $f_1=g_1+1$ and $C_0$ is constant. The last  equation is
 the Bessel one~\cite{Morse} and
$f_1=1+J_0(\rho)$, where $J_0(\rho)$ is zero rank Bessel function.
Finally
  \begin{equation}
  \zeta(\rho,z)=C_0 \left[1+J_0(\rho)\right](e^z-e^{-z})\ ,
  \end{equation}
and according to~(\ref{tilzetd}) it gives an expression for the
function $\tilde\zeta_{,z}$
\begin{equation}
\tilde\zeta_{,z}=C_0J_0(\rho)(e^z+e^{-z})\ ,
\end{equation}
which takes the simple form $\tilde\zeta_{,z|z=0}=2C_0J_0(\rho)$
at the plane  $z=0$. For the large arguments
$\rho\to\infty$, zero rank Bessel
function has the following asymptotic behavior
\begin{equation}
\lim_{\rho\to\infty}J_0(\rho)=\sqrt{\frac{2}{\pi\rho}}\cos(\rho
-\frac{\pi}{4})
\end{equation}
  and consequently
\begin{equation}
\tilde\zeta_{\rm {,z|z=0}}=2C_0\sqrt{\frac{2}{\pi\rho}}\cos(\rho
-\frac{\pi}{4})= \frac{1}{\sqrt{\rho}}\cos(\rho
-\frac{\pi}{4})\equiv x\ ,
\end{equation}
where we put, for the simplification
$C_0=\frac{1}{2}\sqrt{\frac{\pi}{2}}$. Using asymptotic
properties of hyperbolic functions one could write
\begin{eqnarray}
\label{25}
&& \cosh(x)=1+\frac{1}{2}\rho^{-1} \cos^2(\rho
-\frac{\pi}{4})+\frac{1}{24}\rho^{-2}\cos^4(\rho -\frac{\pi}{4})
+O(\rho^{-2})\ ,\\
\nonumber\\
\label{26}
&& \sinh(x)=\rho^{-\frac{1}{2}}\cos(\rho
-\frac{\pi}{4})+ \frac{1}{6}\rho^{-\frac{3}{2}}\cos^3(\rho
-\frac{\pi}{4})+O(\rho^{-2})\ .
\end{eqnarray}

Substituting  obtained asymptotic expressions (\ref{25}) and
(\ref{26}) in solution~(\ref{16}), we can write the asymptotics of function
$f(\rho, z)$, at the plane
 $z=0$ when $\rho\to\infty$, in the following form
\begin{eqnarray}
\label{asympt} \lim_{\rho\to\infty}f_{|z=0}=
\alpha^{-1}\Bigg[1-\frac{1}{4}\rho^{-1}-
\frac{1}{4}\rho^{-1}\sin(2\rho)-\frac{1}{24}\rho^{-2}\cos^4(\rho
-\frac{\pi}{4})+ \nonumber\\ \nonumber\\
+\frac{\beta}{\alpha}\rho^{-\frac{1}{2}}\cos(\rho-\frac{\pi}{4})
+\frac{\beta}{6\alpha}\rho^{-\frac{3}{2}}\cos^3(\rho-\frac{\pi}{4})\Bigg]
+O(\rho^{-2})\ .
\end{eqnarray}
We can see from the asymptotics (\ref{asympt}) that in contrast to
the Papapetrou one (\ref{17}) there is a term being proportional
to $\rho^{-1}$ and at the same time we have asymptotically flat
solution at $z=0$ plane.

\section{General solution}

The general  solution for function $\zeta (\rho, z)$ can be derived by
the simple integration of right hand side of the Poisson like
equation (\ref{pois}):
\begin{equation}
\label{28}
\zeta (\rho,z)=-
\frac{1}{4\pi}\int^{\infty}_{-\infty}\frac{\eta(z')dz'}
{\sqrt{\rho^2+(z-z')^2}}\ .
\end{equation}
Thus in the general case the solution (\ref{16}) for function $f$
contains function $\zeta(\rho,z)$ which satisfy to the integral
(\ref{28}) and as result function $\eta(z)$ generates new
solutions for the function $f(\rho,z)$. At the end we would like
to underline
that for this class of functions $\zeta$, satisfying
to the Poisson like equation
(\ref{pois}), this is most general solution.

Using nonvanishing components of Einstein vacuum equations
\begin{equation}
R_{11}=R_{12}=R_{22}=0
\end{equation}
and following to the book~\cite{Djamal} we can easily obtain
the connection of function $\mu (\rho,z)$ with functions $\zeta
(\rho,z)$, $\omega (\rho,z)=C\rho\zeta_{,\rho}$ and $f(\rho,z)$:
\begin{eqnarray}
&&\mu_{,\rho}=-\frac{f_{,\rho}}{f}+\frac{\rho}{2f^2}
(f^2_{,\rho}-f^2_{,z}) - \frac{f^2}{2\rho}
(\omega^2_{,\rho}-\omega^2_{,z} )\ , \\
\nonumber\\
&&\mu_{,z}=-\frac{f_{,z}}{f}+\frac{\rho}{f^2}f_{,\rho}f_{,z}
 - \frac{f^2}{\rho}\omega_{,\rho}\omega_{,z} \ ,
\end{eqnarray}
and therefore the metric, presented by expression $(1)$  is defined
completely.

 One can see form equation~(\ref{16}) that the general solution
is invariant under the transformation $$\zeta\to\zeta +h(z)$$ in the
functional space of function $\zeta$, where function $h(z)$ satisfies
to the condition $$h(z)_{,zz}=\eta(z).$$ This is gauge
condition directly connected with the
source of gravitational field defined by the metric function $\eta(z)$.
From our point of view this statement is analogous to
the Lorentz gauge condition for the  Maxwell equations in
 electrodynamics.

\section{Spacetime of line gravitomagnetic  monopole}

It is obviously, that any solution of Einstein equations becomes
important if it has any physical application and identified with
the source of gravitational field. Here we demonstrate, that the
obtained extension of Papapetrou solution contains, for example,
solution for the line gravitomagnetic monopole and can be useful
in the interpretation of the connection of the metric source with
the gravitomagnetic monopole momentum $L$.

If we take $\zeta(\rho,z)$ in the form
\begin{equation}
\label{zeta_m} \zeta(\rho,z) = z\ln
\left(\frac{\rho}{\rho_0}\right)^{2m}+\int F(z)dz+\eta_0\ ,
\end{equation}
where $m$, $\rho_0$, $\eta_0$ are some constants and
$F(z)=\int\eta(z)dz$, then as it follows from the
expression~(\ref{zeta_m}), so defined function $\zeta(\rho,z)$
satisfies to the equation $(7)$ for an arbitrary
 function $\eta(z)$. Choosing in our solution $(16)$
$\beta = 0$ and $\alpha = {L}/{2m}$, for the simplicity, we can
obtain
\begin{equation}
\label{f_rho}
f(\rho,z) = \frac{2m}{L}\cdot
\frac{1}{\cosh\left(2m\ln\frac{\rho}{\rho_0}\right)}\ .
\end{equation}
The
function~(\ref{f_rho}) coincides with the metric function
$g_{00}(\rho)$ from the paper~\cite{NZ} and therefore presents the
space time around this source, reproducing cylindrical analogue of
NUT space, named for Newman, Tamburino and Unti~\cite{NUT}.

From this simple example one can see using $(32)$ that the
consideration of the line gravitomagnetic monopole metric in the
Weyl - Lewis -Papapetrou form leads to
\begin{equation}
\label{omega} \omega=\alpha\rho\zeta_{,\rho}=2m\alpha\cdot z=L\cdot z\ ,
\end{equation}
 from where one can conclude, that gravitomagnetic monopole
 momentum $L=2m\alpha$ in accordance with definition of parameter
$\alpha$.
 Further investigation of the effect
 of gravitational field of a line gravitomagnetic monopole
on test electromagnetic fields has been considered in our accompanying
paper~\cite{ban}.

\section{Conclusion}

 We conclude, that obtained solution~(\ref{16}) is an
extension of the well known Papapetrou class of solutions, which
can be applied to the physical systems, presenting rotating
bounded nonzero masses. We have constructed the  solution for the
particular function $\eta(z)$, which in some sense, demonstrates
the existence of $\rho^{-1}$ term in asymptotics. In addition
using properties of Poisson like equation for $\zeta$ function, we
have found the general solution (\ref{28}) for any given function
$\eta (z)$, which generates various metric functions $f(\rho, z)$.
The connection of above obtained solution with metric of a line
gravitomagnetic monopole was considered as a particular case.
This is preliminary result and the physical interpretation of the
obtained solution is outside the framework of this paper. More
detailed investigation and application to gravitational sources is
the scope of our future research. In this respect as the next step
we want to investigate the regularity conditions at the axis
$\rho=0$ in dependence from the choice of function $\eta (z)$.

\section*{Acknowledgments}

Financial support for this work is partially provided by the UzFFR
(project 01-06) and projects F2.1.09 and F2.2.06 of the UzCST.
A.K. and B.A. thank TWAS for the support under the associate
program and IUCAA for warm hospitality during their stay there.
Authors are grateful to anonymous referee for bringing into
attention gauge freedom condition, M. Sami for fruitful
discussions and Mofiz Uddin Ahmed for help and cooperation. A.K.
also thanks Yuri Shtanov for the reference to the
paper~\cite{Fomin1} and useful comments.


\end{document}